\documentclass[twocolumn,prb,twoside,a4paper]{revtex4}
\pdfoutput=1
\renewcommand{\familydefault}{\sfdefault}

\usepackage[font=footnotesize,labelfont=sf,bf,flushleft]{caption}
\usepackage{graphicx}
\usepackage[usenames,dvipsnames]{xcolor}
\usepackage{amsmath,bbm}
\usepackage{natbib}
\usepackage[normalem]{ulem}
\usepackage{titlesec}
\newcommand{\white}[1]{{\color[rgb]{1,1,1}{#1}}}

\titleformat{name=\section}{\filright \sf \normalsize \bfseries}{\thesection}{.5em}{}
\titleformat{name=\subsection}{\filright \sf \small \bfseries}{\thesubsection}{.5em}{}

\newcommand{\shorteq}{%
  \settowidth{\@tempdima}{-}% Width of hyphen
  \resizebox{\@tempdima}{\height}{=}%
}

\begin{document}

\renewcommand{\baselinestretch}{2}
\title{\flushleft{ \Huge{\bfseries Cavity cooling a single charged nanoparticle.}}}

\author{\hspace{-2.2cm} \bf J. Millen\footnotemark[1], P. Z. G. Fonseca, T. Mavrogordatos, T. S. Monteiro \footnotemark[2] \& P. F. Barker\footnotemark[3] }

\affiliation{\white .  \vspace{-1.4cm}}

\maketitle

\renewcommand{\baselinestretch}{1}

\footnotetext[1]{j.millen@ucl.ac.uk}
\footnotetext[2]{t.monteiro@ucl.ac.uk}
\footnotetext[3]{p.barker@ucl.ac.uk}

\small 
% citations form 16 words
\noindent {\bfseries
The development of laser cooling coupled with the ability to trap atoms and ions in electromagnetic fields, has revolutionised atomic and optical physics, leading to the development of atomic clocks, high-resolution spectroscopy and applications in quantum simulation and processing. However, complex systems, such as large molecules and nanoparticles, lack the simple internal resonances required for laser cooling. Here we report on a hybrid scheme that uses the external resonance of an optical cavity, combined with radio frequency (RF) fields, to trap and cool a single charged nanoparticle. An RF Paul trap allows confinement in vacuum, avoiding instabilities that arise from optical fields alone, and crucially actively
 participates in the cooling process. This system offers great promise for cooling and trapping a wide range of complex charged particles with applications in precision
 force sensing \cite{Geraci10}, mass spectrometry, exploration of quantum mechanics at large mass scales \cite{Diosi87, Penrose96} and the possibility of creating large quantum superpositions\cite{Arndt14}.}

\rm

Light is an exceptionally flexible and precise tool for controlling matter, from trapping single atoms \cite{Grangier01} to the optical tweezing of organic cells \cite{Ashkin87}. Laser cooling of atoms has led to the production of the Bose-Einstein condensate \cite{Leggett01} and confinement in optical lattices \cite{Bloch05}, allowing the exploration of quantum physics. The control of nano- and micro-scale systems with light has enabled the study of mesoscopic dynamical and thermodynamical processes \cite{Raizen10, Gieseler14, Millen14}. In recent years it has been possible to control the motion of nanoscale oscillators at the quantum level using light \cite{Lehnert11, Painter11_2}, which is seen to be of fundamental importance for the construction of quantum networks \cite{Cleland13, Lehnert14, Hakonen14}, and has enabled the generation of nonclassical states of light \cite{Painter13}.

\begin{figure}[t]
{\includegraphics[width=0.46\textwidth]{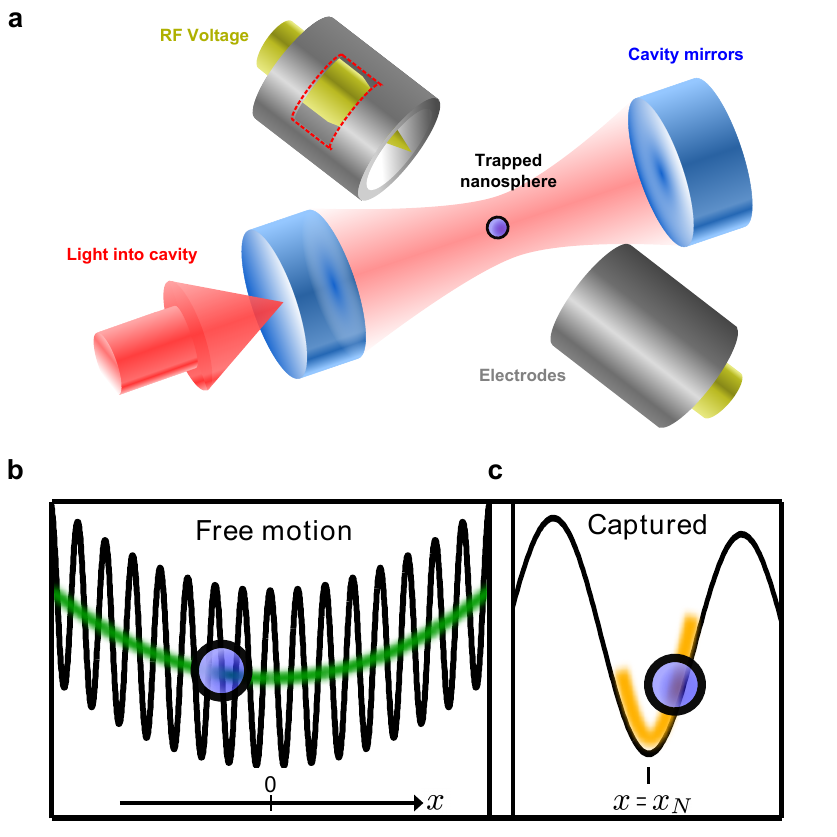}}	
\caption{\label{fig:schem} 
\footnotesize {\bfseries Hybrid electro-optical trap used to levitate and cool a charged silica nanosphere in vacuum.}
{\bfseries a,} A schematic diagram of the hybrid trap, consisting of a Paul trap and standing wave optical cavity potential.
{\bfseries b,} The sphere is trapped by the Paul trap and driven across the standing wave formed by the light in the cavity, scattering light. 
{\bfseries c,} When cooled, the sphere is trapped by the optical field and confined to one fringe of the standing wave. 
}
\end{figure}

The creation of long-lived macroscopic quantum states of nanomechanical oscillators is challenging due to strong coupling with the environment. A mechanical oscillator, such as a nanosphere, that is levitated in vacuum minimizes the coupling to the environment. The lack of clamping leads to extremely high mechanical quality factors \cite{Chang10} ($Q\approx10^{12}$) offering force sensing with exquisite sensitivity \cite{Geraci10}. Active feedback cooling has achieved milli-Kelvin centre-of-mass temperatures \cite{Li11, Gieseler12}. In addition, the ability to change the properties of the oscillator by changing the levitating field, and even to rapidly turn off the levitation, offers the prospect of interferometry in the absence of any perturbations other than gravity \cite{Romero-Isart11}.

Cavity cooling \cite{Horak97} has been used to cool atoms \cite{Vuletic03, Rempi04} and atomic ions \cite{Vuletic09}, and it has been predicted that it could be used to cool larger particles to the centre-of-mass quantum ground-state \cite{Barker10, Chang10, Isart10}. The cooling of neutral nanoparticles transiting a cavity in vacuum \cite{Asenbaum13}, and trapped at 1\,mbar \cite{Kiesel13} has been demonstrated.

We present a new method for cavity cooling charged nanoparticles, utilizing both the optical field of a cavity and the electric field of a Paul trap. Crucially, the Paul trap drives the cavity cooling dynamics by introducing a cyclic displacement of the equilibrium point of the mechanical oscillations in the optical field. In addition the deep Paul trap potential avoids instabilities that have been observed in all-optical traps at low pressures \cite{Gieseler13, Millen14}, and allows us to operate in vacuum.

%Light can be used to cool the motion of nanoparticles, either through active feedback cooling, where milli-Kelvin temperatures are achievable \cite{Li11, Gieseler12}, or through passive cavity sideband cooling \cite{Kiesel13, Asenbaum13}, which offers the potential to reach the centre-of-mass quantum ground-state in the micro-Kelvin range \cite{Barker10, Chang10, Isart10}. However, absorption of light from the intense fields required to simultaneously levitate and cool the nanoparticles \cite{Millen14}, and thermal driving due to collisions with the background gas \cite{Gieseler13} leads to instabilities in the trap at low pressures, meaning to date it has been impossible to simultaneously trap and cavity cool nanoparticles below 1\,mbar. 

%We overcome this problem by levitating a naturally charged silica nanosphere in a hybrid electro-optical trap combining a Paul trap with an optical dipole trap formed from a single mode optical cavity. The electric trap offers deep trapping potentials without instability in high vacuum. Crucially, the Paul trap also plays an important role in in the cavity cooling dynamics by introducing a cyclic displacement of the equilibrium point of the mechanical oscillations in the optical field. This eliminates the need for a second, dedicated cooling optical mode of the cavity \cite{Chang10, Pender12, Monteiro13} and importantly allows us to cavity cool the trapped particle in vacuum.

A schematic diagram of the hybrid electro-optical trap which combines both a Paul trap and an optical cavity dipole trap is shown in fig.~\ref{fig:schem}a.

\begin{figure}[t]
	 {\includegraphics[width=0.46\textwidth]{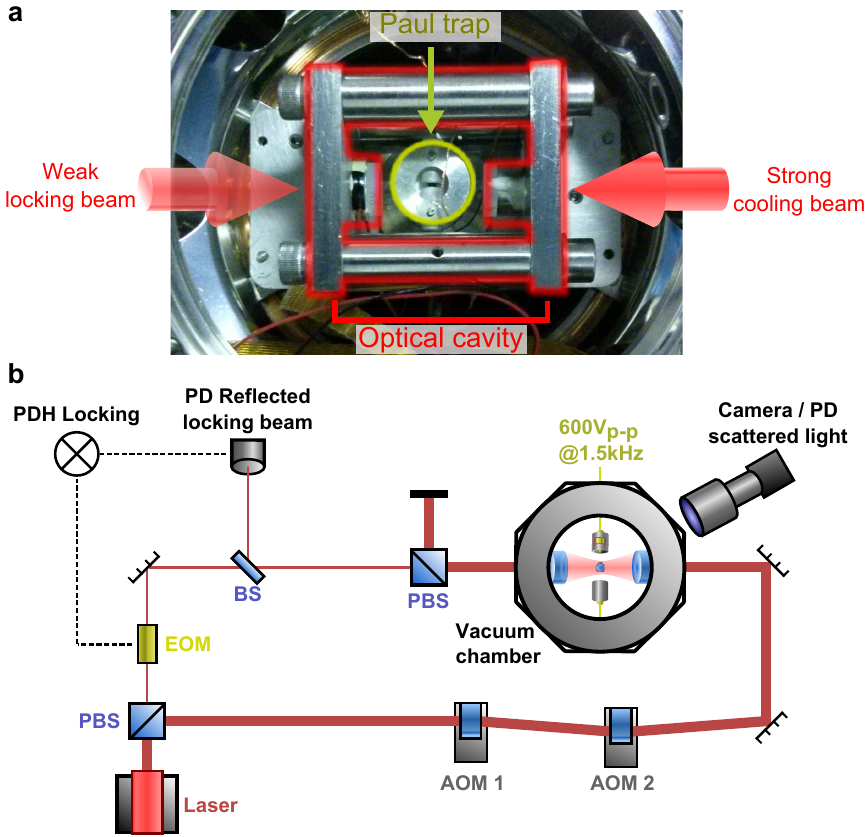}}	
\caption{\label{fig:exp} 
\footnotesize {\bfseries Experimental setup for trapping and cooling nanospheres.} 
{\bfseries a,} A photograph of the experiment. A Paul trap sits in the middle of an optical cavity.
{\bfseries b,} Optical layout for cooling. A weak beam resonant with the optical cavity is used to stabilize the cavity through Pound-Drever-Hall locking, and a beam at least fifty times more intense, which is used for trapping and cooling, can be detuned from the cavity resonance. The light which the nanosphere scatters as it moves through the optical cavity field is collected. 
}
\end{figure}

The Paul trap potential for a nanosphere is approximated by

\begin{equation}
V(x,y,z,t)= \frac{1}{2}m\omega_\mathrm{T}^2 \left(x^2+y^2 -2z^2 \right) \cos (\omega_\mathrm{d} t),
\end{equation}

where $m$ is the mass of the nanosphere and $\omega_\mathrm{T}^2= \frac{2 q V_0}{m r_0^2}$, with $q$ the charge on the nanosphere, and $r_0$ a parameter setting the scale of the trap potential. 

The optical cavity field creates a potential given by $V_{\mathrm{opt}}(x,y,z)= U_0 \cos^2{kx} \ e^{-2(y^2+z^2)/w^2}$
where  the depth of the optical potential $U_0=\hbar A |\alpha|^2$; here $|\alpha|^2$ is the intra-cavity
photon number and the coupling strength  $A = \frac{3V_\mathrm{s}}{2V_\mathrm{m}}\frac{\epsilon_\mathrm{r}-1}{\epsilon_\mathrm{r}+2} \omega_\mathrm{l}$
depends on the sphere volume $V_\mathrm{s}$, the mode volume $V_\mathrm{m}=\pi w^2L$ (with $w$ the waist of the cavity field and $L$ the length of the cavity) and the laser frequency $\omega_\mathrm{l}$. The optical potential gives rise to a mechanical frequency $\omega_\mathrm{M}$, which in the experiments presented in this paper is $\omega_\mathrm{M}\simeq 2\pi \times 18$\,kHz.

We define two regimes of motion within the hybrid trap. Free motion refers to when the particle is driven across the optical standing wave potential by the Paul trap (fig.~\ref{fig:schem}b), while captured motion corresponds to where the nanosphere is confined and oscillates within a single well of the standing wave (fig.~\ref{fig:schem}c), where most of the cooling takes place.

\begin{figure}[t]
	 {\includegraphics[width=0.46\textwidth]{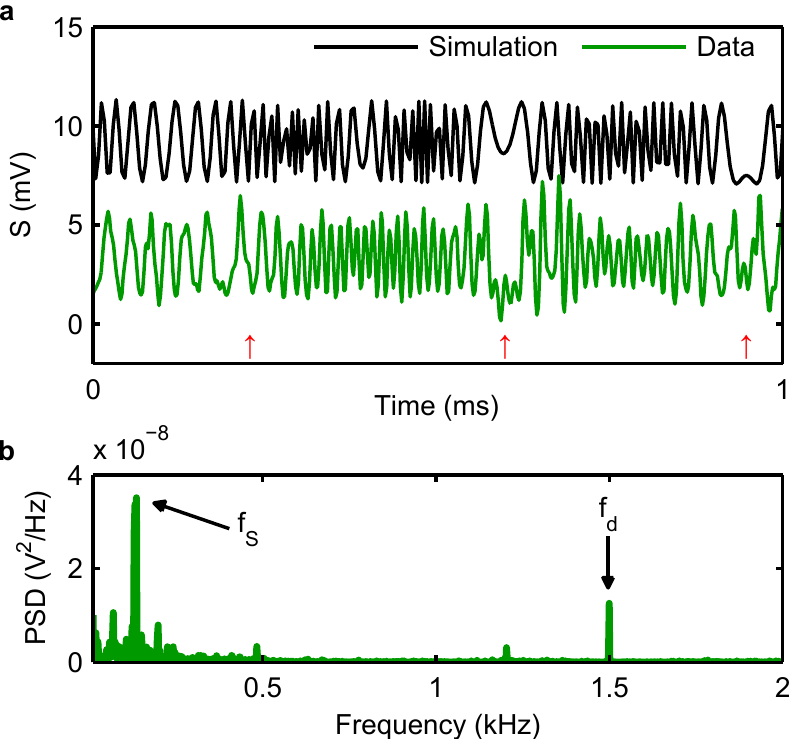}}	
\caption{\label{fig:free} 
\footnotesize {\bfseries Free motion of the nanosphere in the Paul trap.} 
{\bfseries a,}  The scattered signal S as the Paul trap drives the nanosphere across the optical standing wave, with a simulation showing the predicted dynamics. The turning points of the motion are marked.
{\bfseries b,} During free motion the power spectral density (PSD) of the nanosphere's motion shows the Paul trap secular frequency $\omega_\mathrm{S} = 2\pi\times \mathrm{f}_\mathrm{S}$ and micromotion $\omega_\mathrm{d} = 2\pi\times \mathrm{f}_\mathrm{d}$. Measurement of $\mathrm{f}_\mathrm{S}$ allows readout of the charge, in this case just a single charge $q/e = 1$.
}
\end{figure}

Near turning points of the Paul trap motion (where the speed of the nanoparticle is a minimum) 
a small amount of damping can lead to the nanosphere losing enough energy to be captured by the
 optical potential within a well at $x = x_N$ (fig.~\ref{fig:schem}c). Optomechanical damping of the motion of the nanosphere occurs when the light is red detuned with respect to the cavity resonance. The damping occurs as the nanosphere oscillates across the optical potential well in the antinode of the light field. This motion modulates the effective cavity length and periodically brings the cavity closer to resonance, increasing the intracavity power and thus the optical potential.  Due to the decay time of light in the cavity the intracavity build-up is delayed with respect to the motion, which causes damping \cite{Barker10}. In the absence of the Paul trap field, the nanosphere is cooled by the light towards the bottom of a local standing wave potential well (optical antinode). In this region, cooling is ineffective as the nanosphere's motion does not significantly modulate the cavity length.  However, in our hybrid system, the additional force on the charged nanosphere by the Paul trap acts to periodically pull the particle away from the anti-node, leading to enhanced cooling.   The oscillation within the optical potential is therefore asymmetric around the bottom of the potential well as illustrated in fig.~\ref{fig:schem}c.

To explore the cooling and particle dynamics in the hybrid trap, we use a Paul trap formed by two rounded electrodes separated by 1\,mm. This traps nanospheres of radius 209 nm which typically have 1-3 elementary charges. An RF voltage of $V_0\cos(\omega_\mathrm{d}t)$, where $V_0$ is the amplitude of the time-varying signal ranges from $300 - 900$\,V and $\omega_\mathrm{d} = 2\pi\times1500$\,Hz is its frequency, is applied to the electrodes. The electrodes are enclosed by grounded cylindrical shields. The intra-cavity field of a medium finesse ($\mathcal{F} \simeq 15,000$) optical cavity passes through the middle of the Paul trap. The hybrid trap was placed within a vacuum chamber as shown in fig.~\ref{fig:exp}a. The nanospheres can be trapped at all pressures down to the limit of $10^{-6}$\,mbar in the current setup. The laser used to provide the trapping and cooling light is split into a strong and weak beam. The weak beam is used to lock the laser to the cavity via the Pound-Drever-Hall method, as illustrated in fig.~\ref{fig:exp}b. The strong beam is used for cooling and trapping and can be arbitrarily shifted in frequency with respect to the cavity resonance. 

\begin{figure}[t]
	 {\includegraphics[width=0.46\textwidth]{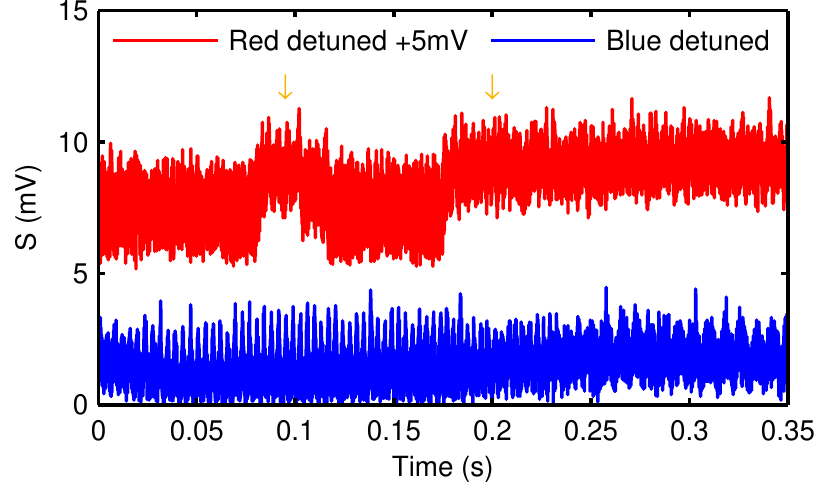}}	
\caption{\label{fig:detuning} 
\footnotesize {\bfseries Effect of detuning from the cavity resonance on the dynamics of the nanosphere.} 
Scattered light signal S from the nanosphere at a pressure of $7.4\times10^{-3}$\,mbar with $\Delta \simeq \pm\kappa$. When the light is red detuned from the cavity resonance (upper red trace, signal offset for clarity) there are periods of free motion in the Paul trap, and regions where the nanosphere is captured in the optical potential, where the particle can be cooled (marked at $t \simeq 0.2$\,s) or not (marked at $t \simeq 0.1$\,s) depending upon the location within the optical standing wave. The nanosphere is not captured by the optical potential when the light is blue detuned from the cavity resonance (lower blue trace) and is periodically heated out of the optical field. 
}
\end{figure}

To observe the dynamics and characterize the cooling of a single particle, we measure the scattered light S as the nanosphere passes through the standing wave field. This provides a measurement of particle position as a function of time. Free motion of the particle produces a rapid modulation as the particle moves through many antinodes of the standing wave cavity field, as shown in fig.~\ref{fig:free}a. During free motion, the trajectory of the nanoparticle is well parametrized by
 the Paul trap secular $\omega_\mathrm{S} \simeq \frac{\omega_\mathrm{T}^2}{\sqrt{2}\omega_\mathrm{d}}$ and micromotion frequency $\omega_\mathrm{d}$ (see fig.~\ref{fig:free}b), despite the addition of the optical field. 

However, when a particle is  captured and cooled it oscillates around a single antinode and the time averaged scattered light increases.  A time series recorded for both free and captured motion is shown in fig.~\ref{fig:detuning}, for two detunings from the cavity resonance. The red trace is taken for light that is red detuned and shows regions of increased intensity as the particle is captured in the standing wave potential. When captured the nanosphere is cooled, with greater cooling rates for captures far from the centre of the Paul trap. A similar trace for blue detuned light shows no capture (a constant mean scattered light intensity), as this detuning leads to heating of the particle motion.

\begin{figure}[t]
	 {\includegraphics[width=0.46\textwidth]{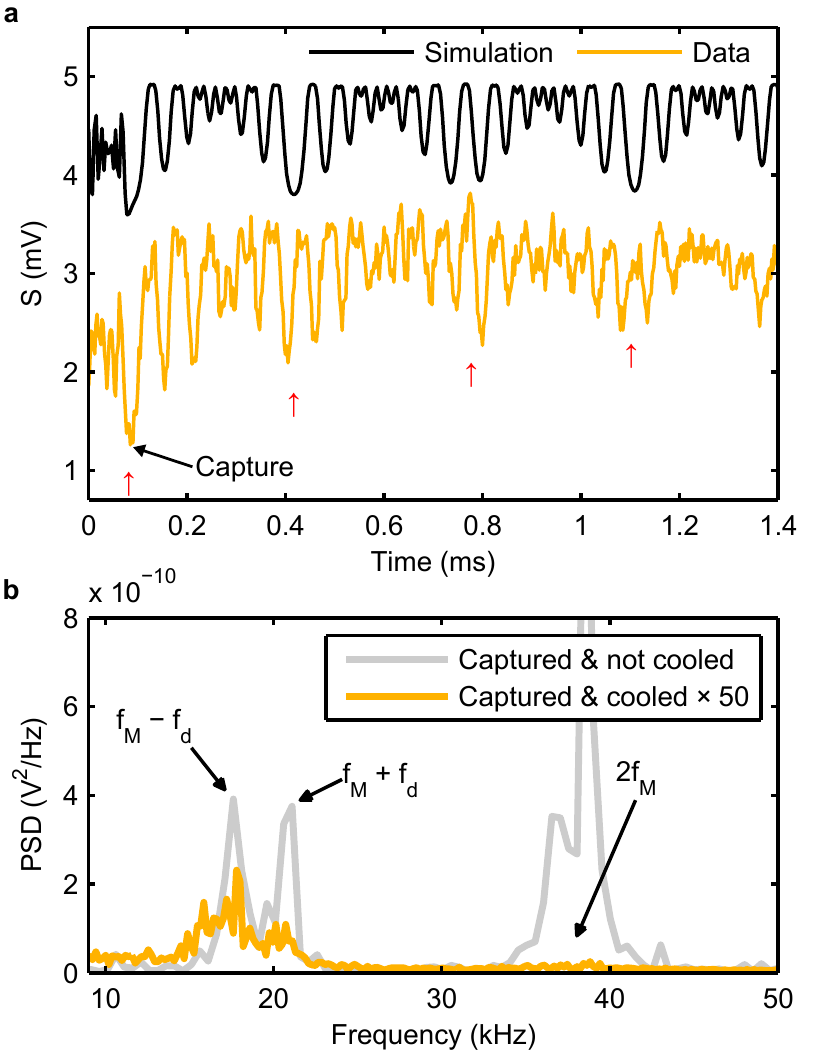}}	
\caption{\label{fig:captured} 
\footnotesize {\bfseries Capture and optomechanical cooling of the nanosphere.} 
{\bfseries a,}  The scattered signal S once the nanosphere has been captured, at a pressure of $4.6 \times 10^{-4}$\,mbar, and a simulation of the process. The mechanical motion is cooled, and the motion due to the Paul trap is still present (marked by arrows).
{\bfseries b,} The PSD of the nanosphere's motion while captured by the optical field. At a low pressure of $4.6 \times 10^{-4}$\,mbar (orange line) the split mechanical frequency $\mathrm{f}_\mathrm{M} \pm \mathrm{f}_\mathrm{d}$ is visible, but the doubled frequency $2\mathrm{f}_\mathrm{M}$ is heavily suppressed, indicating cooling. At a higher pressure of $7.4\times10^{-3}$\,mbar (grey line) mechanical damping from the background gas dominates and the nanoparticle is not cooled, as shown by the unsuppressed peak at $2\mathrm{f}_\mathrm{M}$.
}
\end{figure}

Figure~\ref{fig:captured}a shows the the capture and initial cooling process. In this trace the initial rapid modulation is followed by a dip in intensity as the particle is trapped. The particle then starts to undergo a well defined motion in the optical well, showing period doubling as the Paul trap causes asymmetric oscillations about the optical antinode (as illustrated in fig.~\ref{fig:schem}c). The period doubling is clear evidence that the Paul trap is modifying the dynamics within the optical potential, which is essential for efficient cooling. The average scattered light intensity increases as the nanosphere is cooled towards the optical antinode.

To quantify the optomechanical cooling we consider the motion of the nanosphere within one well of the optical potential $x^{\prime}(t)=x(t)-x_N$. Within the optical potential a mechanical frequency $\omega_\mathrm{M}$ can be defined, which depends upon $U_0$ as well as which well $N$ of the standing wave the particle is confined to (see supplementary material for full details). Since $\omega_\mathrm{M}\gg \omega_\mathrm{S},\omega_\mathrm{d}$ the motions due to the optical and Paul trap potentials are adiabatically separable; as $t \to \infty$, we approximate the motion along the optical axis by:

\begin{equation}
x^{\prime}(t) \sim X_\mathrm{d}\cos(\omega_\mathrm{d} t) + X_M^{\infty}\cos(\omega_\mathrm{M} t).
\end{equation}

The drive amplitude 
$X_\mathrm{d} \approx \frac{\omega_\mathrm{T}^2}{\omega_\mathrm{M}^2}  x_N$
is largely undamped. $X_\mathrm{M}^{\infty}$ is the steady-state amplitude of the 
mechanical motion; initially, $X_\mathrm{M}(t)$ decays as $
 \sim e^{-\Gamma_{\mathrm{opt}}t}$ but tends to a steady state 
value determined by noise heating processes.
 By analyzing the Fourier spectrum of S due to $x^{\prime}(t)$ one finds, to leading order,
 frequencies at $\omega_\mathrm{M} \pm \omega_\mathrm{d}$ with amplitude $A_1$ and a single
 peak at $2\omega_\mathrm{M}$ with an amplitude $A_2$ (see supplementary material). 
The amplitudes of the Fourier peaks $A_{1,2}$ are related to the amplitude of the motion of the nanosphere through
 $\frac{A_2}{A_1} \simeq  \frac{J_0(2kX_\mathrm{d})J_2(2kX_\mathrm{M}^{\infty})}{J_1(2kX_\mathrm{d})J_1(2kX_\mathrm{M}^{\infty})}$ where $J_n$ are Bessel functions of order $n$ and
 $\frac{A_2}{A_1} \to \frac{X_\mathrm{M}^{\infty}}{X_\mathrm{d}}$ for small amplitude oscillations. Hence, by analyzing the Fourier spectrum the cooling rate of the nanoparticle's motion can be measured.

In fig.~\ref{fig:captured}b, the Fourier spectrum of captured nanosphere motion is shown. 
At a pressure of $7.4\times10^{-3}$\,mbar (grey line) the damping due to the background gas $\Gamma_\mathrm{M} \sim \Gamma_{\mathrm{opt}}$, hence no cooling is evident and $A_2 > A_1$. At a pressure an order of magnitude lower, where $\Gamma_{\mathrm{opt}} \gg \Gamma_\mathrm{M}$ (orange line), both the capture and subsequent cooling are dominated by optomechanical damping and there is a significant reduction in the size of $A_2$. The width of peaks is due to the excursion of the nanosphere into the anharmonic regions of the optical potential.

We can also analyze the cooling by calculating the optomechanical damping rates directly. 
For small oscillations, the cooling of a levitated particle
may be obtained from the well-known optomechanical quantum Hamiltonian:

\begin{equation}
 \frac{{\hat H_{\mathrm{Lin}}}}{\hbar} =  -\Delta {\hat a}^\dagger {\hat a} 
  + \frac{\omega_\mathrm{M}}{2}({\hat x}^2+{\hat p}^2) + g ( {\hat a}+  {\hat a}^\dagger) {\hat x}
\end{equation}

where ${\hat x}$ is a small fluctuation (in units of $X_{\mathrm{ZPF}}=\sqrt{\hbar/2m\omega_\mathrm{M}}$)
 about an equilibrium point $x_0$ and ${\hat a}$ is a fluctuation of optical field the about the mean 
value $\alpha$.
Whether in the quantum regime or not, the corresponding damping $\Gamma_{\mathrm{opt}}$ is then given by
$\Gamma_{\mathrm{opt}}= g^2 \kappa\left[\mathcal{S}(\omega_\mathrm{M})-\mathcal{S}(-\omega_\mathrm{M})\right]$ 
where $\mathcal{S}(\omega)= \left[[\Delta-\omega]^2 +\frac{\kappa^2}{4}\right]^{-1}$.
Simulations with the full 3D nonlinear dynamics of the hybrid trap shows that the above expressions
are accurate provided one allows for the slow
excursion in the equilibrium point, since $x_0 \equiv x_0(t) \simeq X_\mathrm{d} \cos \omega_\mathrm{d} t$ and the equilibrium point can be far from the antinode ($0 \leq kX_\mathrm{d}  \lesssim \pi/2$). This consideration leads to an effective optomechanical coupling (see supplementary information):

\begin{eqnarray}
 g^2= \frac{\hbar k^2 A^2 |\alpha|^2}{2m  \omega_\mathrm{M}} \left(1-J_0(4kX_\mathrm{d})\right).
\label{LinHam}
\end{eqnarray}

Using the experimental parameters for the data in fig.~\ref{fig:captured}a ($\mathcal{F} \simeq 15000$,
$\Delta=2\pi \times 288\,$kHz, intracavity power $P = 14\,$W, $\omega_\mathrm{S}=2\pi \times 100\,$Hz) we find
$\Gamma_{\mathrm{opt}}\approx 20-30$ Hz, where the capture is at well position $N=kx_N/\pi  \simeq 600-700$. The expected equilibrium temperature
$T_{\mathrm{eq}} \simeq \frac{\Gamma_\mathrm{M}}{ \Gamma_{\mathrm{opt}}} T_\mathrm{B}$, where here $T_\mathrm{B}=300$K, and $\Gamma_\mathrm{M} \simeq 1$ Hz
 at $4.6 \times 10^{-4}$ mbar pressure. This means that the motion of the nanosphere is cooled to $\sim 10\,$K, which represents a factor of $\sim 100$ reduction in energy from the well depth $U_0$. This reduction in energy is consistent with the Fourier analysis of the peaks in fig.~\ref{fig:captured}b.

The nanospheres are confined within an optical well for a time of order 0.1-0.2s, far longer than the time needed to attain a steady state temperature (of order 10 ms). After this time they escape the optical potential since both the optomechanical coupling $g$ and the optical potential $U_0$ are very weak, and noise on the light field is sufficient to kick the nanospheres out of the well, a situation we have confirmed with simulations. Once the spheres are lost from the optical well they are recaptured by the Paul trap and returned to the cooling cycle. To solve the problem of loss from the optical potential a higher finesse optical cavity can be used, with a smaller mode volume $V_\mathrm{m}$, since $g^2 \propto V_\mathrm{m}^{-2}$ and we can operate at the sideband-resolved regime
$-\Delta \simeq \omega_\mathrm{M}  \gtrsim \kappa/2$ which yields maximum cooling, and size of the optical potential $U_0 \propto V_\mathrm{m}^{-1}$ is enhanced. For example, for a cavity with $\mathcal{F} = 100,000$ and $w = 60\,\mu$m cooling rates of $\sim 10\,$kHz are possible.

We have reached low temperatures with very low cooling rates of 30\,Hz, and the introduction of a higher finesse optical cavity would greatly lower our final temperature. So far we have only discussed cooling in the $x$ direction, along the optical axis. Our simulations show that cooling in the $y$ and $z$ directions can be achieved with the addition of a small angle between the coordinate systems of the optical and Paul trap fields. 

In conclusion, we present a hybrid optical and quadrupole electric trap for both confining and optomechanically cooling charged levitated dielectric objects in high vacuum. Since the levitation is not by an optical field alone, light intensities that have been shown to melt silica nanospheres \cite{Millen14} do not have to be used. The system we present in this paper is very suitable for cooling nanoscale objects to the quantum level, in principle the only required modification is a higher finesse optical cavity. This work opens the way for the exploration of fundamental questions in quantum physics, and the development of macroscopic quantum technologies. Since Paul traps are used widely for the characterisation of nanoparticles, our work introduces a powerful new tool for their manipulation and the damping of their motion.

\section*{Acknowledgements}

\noindent We acknowledge support from the UK's Engineering and Physical Science Research Council grant EP/H050434/1.

\section*{Methods}

Silica nanospheres of radius 209\,nm (Corpuscular inc.), were sonicated in methanol for several hours to prevent clumping. The solution was dropped onto a ceramic piezoelectric speaker and allowed to dry. Nanospheres were launched into the trap from the speaker through a narrow aperture into the Paul trap, which is loaded at a pressure of $2\times 10^{-1}\,$mbar. During loading the cavity mirrors are covered by a shield to protect them from the nanoparticles. However, the finesse of the cavity varied between $\mathcal{F} = 10,500 - 15,000$ over the course of the experiments due to contamination of the cavity mirrors by nanoparticles. The charge of the nanospheres could be ascertained from the Paul trap secular frequency $\omega_\mathrm{S}$.
 
The Paul trap is mounted on a remotely-controllable, vacuum compatible, $x-y-z$ translation stage which was used to align the trapped particle with the optical axis of the cavity. The cavity has a length of 3.7\,cm a beam waist of 140\,$\mu$m.

\begin{center}
\textsc{\LARGE Supplementary Information}\\[1.2cm]
\end{center}

\section{Dynamics in a Paul trap}

The dynamics and stability properties of quadrupole ion traps have been extensively studied,
 given their importance for scientific research as well as applications such as ion 
 mass spectrometry see (see  \cite{March} for a review), so here we simply outline some key features.
A general Paul trap combines a DC static field with an AC component, oscillating with a drive 
frequency $\omega_{\textrm{d}}$. The corresponding potential is
 $V_{\textrm{ion}}(x,y,z,\omega t)= V^{\textrm{AC}}(x,y,z,\omega t) +V^{\textrm{DC}}(x,y,z)$
 with oscillating component:

\begin{eqnarray}
V^{\textrm{AC}}(x,y,z,t)= \phi^{\textrm{AC}}_0 \left(\beta_{\textrm{x}} x^2+\beta_{\textrm{y}} y^2+ \beta_{\textrm{z}} z^2 \right) \cos (\omega_{\textrm{d}} t),
\end{eqnarray}

\noindent and a DC field of a similar form:

\begin{equation}
V^{\textrm{DC}}(x,y,z)= -\phi^{\textrm{DC}}_0 \left(\beta_{\textrm{x}} x^2+\beta_{\textrm{y}} y^2+ \beta_{\textrm{z}} z^2 \right),
\end{equation}

\noindent where $\phi^{\textrm{AC(DC)}}_0= Q V_0(U_0)/r_0^2$. Here, $Q$ is the charge of the particle, $V_0(U_0)$ is the applied AC(DC) voltage and the parameter $r_0$ sets the scale of the Paul trap potential. 
For a 3D quadrupolar trap, $\beta_{\textrm{x}}=\beta_{\textrm{y}}=1$ while $\beta_{\textrm{z}}=-2 $, constraints imposed by the requirement that the potential field obey Laplace's equation. 
The resulting motion is separable; taking ${\textrm{u}} \equiv x,y,z$, the force $F_{\textrm{u}}=-\partial V/\partial u$ 
yields three equations of motion, one for each degree of freedom:

\begin{equation}
m{\ddot {\textrm{u}}}=\frac{m \omega_{\textrm{d}}^2}{4}[a_{\textrm{u}}-2q_{\textrm{u}}\cos(\omega_{\textrm{d}} t)]{\textrm{u}},
\label{Mathieu}
\end{equation}

\noindent where the DC and AC stability parameters are respectively $a_{\textrm{x,y}}=Q U_0/(m \omega_{\textrm{d}}^2 r_0^2)$ and $q_{\textrm{x,y}}=4Q V_0/(m \omega_{\textrm{d}}^2 r_0^2)$, while $a_{\textrm{z}}=-a_{\textrm{x}}$ and $q_{\textrm{z}}=-2q_{\textrm{x}}$.
Eq.\ref{Mathieu} represents three equations in Mathieu form; these have been
shown to be stable for specific ranges of  $a_{\textrm{u}},q_{\textrm{u}}$.
 In our experiment, we consider a Paul trap with zero applied DC voltage ($U_0,a_{\textrm{u}}=0$). In this case, the 
ion trap motion becomes unstable for $ q_{\textrm{u}} \gtrsim 0.91$.
Conversely, for small enough $q_{\textrm{u}} \lesssim 0.2$, the motion is not only stable but
may also be adiabatically separated into a 
fast micromotion and a slower secular motion; the corresponding trajectories are given by:

\begin{equation}
{\textrm{u}}(t) \simeq {\textrm{u}}_0 \cos(\omega_{\textrm{S}} t)-{\textrm{u}}_0\frac{q_{\textrm{u}}}{2}\cos(\omega_{\textrm{S}} t)\cos(\omega_{\textrm{d}} t),
\end{equation}

\noindent where
 $\displaystyle\omega^{({\textrm{u}})}_{\textrm{S}} \simeq \frac{\omega_{\textrm{d}}}{2}\sqrt{a_{\textrm{u}}+\frac{q_{\textrm{u}}^2}{2}}$ is
 the secular frequency and ${\textrm{u}}_0$ an initial amplitude.
 In the main text we set $ \omega^{({\textrm{x}})}_{\textrm{S}}=\omega^{({\textrm{y}})}_{\textrm{S}} \equiv \omega_{\textrm{S}}$
 noting that the equality signs strictly apply only to a perfectly quadrupolar trap and in the absence of 
perturbations arising from the cavity field (see below).
We have redefined the ion trap potential in terms of a characteristic 
ion trap frequency $\omega^2_{\textrm{T}}=2Q V_0/(mr_0^2)$ which is typically of the same order
as $\omega_{\textrm{S}}$.  In the separable regime,
$\omega^2_{\textrm{T}} \simeq \sqrt{2} \omega_{\textrm{d}}\omega_{\textrm{S}}$.

\section{Equations of motion in the hybrid trap}

The combined dynamics of the hybrid cavity and Paul trap potentials requires the
 solution of coupled equations of motion for the mechanical degrees of freedom $\hat{x},\hat{y},\hat{z}$ 
 as well as the cavity light mode $\hat{a}$. 
The equations of motion are obtained here from the total Hamiltonian of the hybrid trap.
In a frame rotating at the laser frequency $\omega_l$, the corresponding quantum
Hamiltonian may be written:

\begin{equation}\label{Hamiltonian}
\hat{H}=\frac{\hat{P}^2}{2m}-\hbar\Delta \hat{a}^{\dagger}\hat{a}
 +\hbar\mathcal{E}(\hat{a}^{\dagger}+\hat{a})-V_{\textrm{opt}}(\hat{x},\hat{y},\hat{z}) +V^{\textrm{AC}}(\hat{x},\hat{y},\hat{z},t).
\end{equation}

The first term is the total kinetic energy operator of the nanoparticle; $\Delta=\omega_l-\omega_c$ represents the detuning
between the laser frequency and the resonant frequency of the cavity $\omega_c$. If $\omega_l < \omega_c$,
the cavity is red-detuned. $\mathcal{E}$ represents the driving amplitude of the cavity and is related to the
input power $P_{\textrm{in}}$ (for a critically coupled cavity) by 
$\displaystyle\mathcal{E}^2=\frac{\kappa}{2}\cdot\frac{P_{\textrm{in}}}{2\hbar \omega_l}$. 
In the experimental regimes under consideration it is reasonable to assume semiclassical behaviour
 and replace the operators by scalar complex amplitudes $ \hat{a}(t) \to \langle \hat{a}(t) \rangle \equiv a(t)$; we write
 the optical potential as follows:
\begin{equation}\label{VOPT1}
V_{\textrm{opt}}(x,y,z)=\hbar A |a(t)|^2\cos^2(k x)\mathcal{F}(y,z)
\end{equation}
where the transverse envelope of the light beam is $\mathcal{F}(y,z)=\exp[-2(y^2+z^2)/w^2]$.
The Paul trap potential is
$V^{\textrm{AC}} = \frac{1}{2} m \omega^2_{\textrm{T}} \left( x^2+ y^2-2 z^2 \right) \cos (\omega_{\textrm{d}} t)$.
Hence, we obtain equations of motion for
 the axial degree of freedom (along the optical axis):

\begin{equation}
\ddot {x}=-\frac{\hbar k A}{m}|a(t)|^2 \sin(2kx)\mathcal{F}(y,z)-\Gamma_{\textrm{M}}\dot{x}-\omega^2_{\textrm{T}}x\cos (\omega_{\textrm{d}} t),
\label{axial}
\end{equation}

\noindent the transverse motions:

\begin{equation}
\begin{aligned}
&\ddot{y}=-\frac{4y}{w^2} \frac{\hbar A}{m} |a(t)|^2\cos^2 (k{x})\mathcal{F}(y,z)-\Gamma_{\textrm{M}}\dot{y}-\omega^2_{\textrm{T}} y\cos (\omega_{\textrm{d}} t),\\
&\ddot{z}=-\frac{4z}{w^2} \frac{\hbar A}{m} |a(t)|^2 \cos^2 (k{x})\mathcal{F}(y,z)-\Gamma_{\textrm{M}}\dot{z} +2\omega^2_{\textrm{T}} z\cos (\omega_{\textrm{d}} t). 
\end{aligned}
\end{equation}

\noindent and for the cavity field:

\begin{equation}
\dot {a}= i\Delta a -i\mathcal{E} +i A a \cos^2(kx)\mathcal{F}(y,z)-\frac{\kappa}{2} a.
\end{equation}
where we have added terms to account for the mechanical damping (of rate $\Gamma_{\textrm{M}}$) due 
to collisions with residual background gas and optical decay terms 
(of rate $\kappa/2$) due to losses at the cavity mirrors.
The resulting set of coupled nonlinear equations were solved to simulate 
the free motion, capture process as well as trapping and cooling.
In the trapped and cooling regimes however, the optical potential
is dominant and oscillations are small. Thus we can analyse the motion by
linearising about equilibrium values.

\section{Optically trapped particle}

For a particle which has been trapped at the $N$-th optical well,
 it is most convenient to switch to a local coordinate frame with
the origin at the antinode, $x=x_N$, of the trapping well;
hence, we transform the axial coordinates to $x'=x-x_N$ and obtain
for the axial motion: 

\begin{equation}\label{fulleq}
\ddot{x}^{\prime}=-\frac{\hbar k A}{m}|a(t)|^2 \sin(2kx')\mathcal{F}(y,z)-
\Gamma_{\textrm{M}}\dot{x}^{\prime}-\omega^2_{\textrm{T}}(x^{\prime}+x_N)\cos(\omega_{\textrm{d}}t).
\end{equation}

For optically levitated particles, consideration of small oscillations about a point
$x'=x_0$ determines a mechanical frequency 
$\omega_M$  where typically $\omega_{\textrm{M}} \gg \omega_{\textrm{d}},\;\omega_{\textrm{S}}$.
Thus we assume that the fast oscillations near the antinode are adiabatically separable from the
slower timescales associated with the quadrupole trap. 
We write $x^{\prime}(t)=x_0(t)+x(t)$, where we assume that the amplitude of the
(fast varying) $x(t)$ is small with respect to the slower oscillations $x_0(t)$.
In the first instance, we assume a quasistatic $x_0(t) \simeq x_0$ and obtain
for the axial motion:

\begin{equation}\label{orders}
\ddot{x_0}+\ddot{x}=-\frac{\hbar k A}{m}|a(t)|^2 \cos(2kx_0)[\sin(2kx)+\tan(2kx_0)\cos(2kx)]-\omega^2_{\textrm{T}}x_N\cos(\omega_{\textrm{d}} t),
\end{equation}
\noindent where we neglect damping due to the presence of the gas (see end of the section)
 and assume that the particle is caught in a well such that $x_N\gg x^\prime$. 
 
Similarly, if we assume temporal fluctuations in the cavity field are
small, we can also transform to shifted values for the field amplitudes hence
$a(t)\to\bar{\alpha}+a(t)$ where $\bar{\alpha}^2$ represents
the mean cavity photon number and $a(t)$ are now small fluctuations about the mean.

Keeping only zero-th order terms in equation \eqref{orders} and assuming that $|\ddot{x}_0|\ll|\ddot{x}|$ we obtain:

\begin{equation}
\omega_{\textrm{M}}^2\tan(2kx_0) \approx -2kx_N\omega^2_{\textrm{T}}\cos(\omega_{\textrm{d}} t),
\end{equation}

\noindent where we write $\omega_{\textrm{M}}^2 \simeq (2\hbar k^2 A \mathcal{F}(y,z)|\bar{\alpha}|^2/m) \cos(2kx_0)$
(and we show below that $\omega_{\textrm{M}}$ is in fact the mechanical frequency
of our system).

 Recalling that $x_0(t)$ is time-dependent and is oscillating (albeit slowly), the above can be recast in the form:

\begin{equation}\label{fullx0}
\sin(2kx_0(t)) \approx -\frac{\omega^2_{\textrm{T}}}{\omega_{\textrm{M}}^2}(2kx_N)\cos(\omega_{\textrm{d}}t).
\end{equation}

\noindent Hence if $x_0$ remains small for all $t$
then $\displaystyle x_0(t) \approx -\frac{\omega^2_{\textrm{T}}}{\omega_{\textrm{M}}^2} x_N \cos(\omega_{\textrm{d}} t)$;
however, we do not need to assume this: the excursion in $x_0$ can in fact span
a significant fraction of the well; in that case, the mechanical frequency $\omega_{\textrm{M}}(t)$ varies
 within the range $\omega_{\textrm{M}}[1-\sqrt{\cos(2kX_{\textrm{d}})}]$, which is approximately $15\%$ of $\omega_{\textrm{M}}$ 
for an oscillation about a distance of $\lambda/4$ from the bottom of the optical potential well. 
This variability in $\omega_{\textrm{M}}$ leads to broadening of the peaks in Fig.(3)
of the main text. For simplicity, we assume in the analysis below (but not in our numerical simulations)
 that the transverse excursions are small
and thus $\mathcal{F}(y,z) \simeq 1$. However, in fact these excursions remain detectable and
 even for the trapped particle,
the scattered light detects oscillations at one or both
transverse secular frequencies $\omega^{(y,z)}_{\textrm{S}}= [\omega_{\textrm{d}}/(2\sqrt{2})]q_{\textrm{y,z}}$. The
transverse oscillations are important
for sensing the full 3-dimensional particle trajectories;
 as they are only weakly affected by the optical potential
they provide a reliable measurement of the nanoparticle's charge, $Q$.
If the transverse motions become significant, they lead to further variability in $\omega_{\textrm{M}}$ and
more importantly, to escape from the optical well.\\

The zero-th order terms in the equation for the optical field yield a value for the mean intracavity fields:
\begin{equation}
|\bar{\alpha}|^2 =\frac{\mathcal{E}^2}{\kappa^2/4 + \Delta^2}
\end{equation}
where $\Delta$ includes a small correction from the optical field \cite{Monteiro} which is neglected here
as the coupling to the optical field $A$ is not very strong.

 The optomechanical coupling which leads to optical damping
 is contained in the first order equations for $x(t)$: 

 \begin{equation}\label{xcool}
\ddot{x} = -\omega_{\textrm{M}}^2 x - \frac{\hbar k A}{m} \bar{\alpha} ({a}+{a}^{*})\sin(2kx_0) -\Gamma_{\textrm{M}}{\dot{x}}
\end{equation}
where, without loss of generality, we have assumed that $\bar{\alpha}$ is real.
(eg in the Hamiltonian, by means of the transformation $\bar{\alpha}\to\bar{\alpha} e^{-i\theta}$ 
and $\hat{a}\to \hat{a} e^{i\theta}$ we can restrict ourselves to real $\bar{\alpha}$) .

For $a(t)$ we obtain:
\begin{equation}\label{acool}
\dot{a} = i\Delta a - ikA \bar{\alpha} x \sin(2kx_0) -\frac{\kappa}{2} a.
\end{equation}

For red-detuning ($\Delta < 0$) the $x(t)$ motion is damped such that
 $x(t) \simeq  X_{\textrm{M}}  e^{-\Gamma_{\textrm{opt}}t} \cos (\omega_{\textrm{M}} t)$
where $X_{\textrm{M}}$ is the initial amplitude for mechanical oscillations (soon after capture by an optical well).
Thus the mechanical motion cools until a steady state is achieved whereby heating rates due to 
residual gas collisions or technical noise equals the cooling rate (see \cite{Monteiro} for a 
study in levitated systems) and hence $X_{\textrm{M}}  e^{-\Gamma_{\textrm{opt}}t} \to X_{\textrm{M}}^\infty$ where $X_{\textrm{M}}^\infty$ is
a steady state amplitude which may fluctuate in time due to background noise; in effect  
 $\langle(X_{\textrm{M}}^\infty)^2\rangle$ represents a good estimate of the variance of the mechanical motion and hence provides
 an estimate of the effective temperature of our system. 

Finally, we combine the mechanical oscillations with the slower excursion in the equilibrium point
to model the axial motion of the captured nanoparticle:
\begin{equation}
x^{\prime}(t) \approx X_{\textrm{d}}\cos(\omega_{\textrm{d}} t) + X_{\textrm{M}}^{\infty}\cos(\omega_{\textrm{M}} t).
\label{ansatz}
\end{equation}

\section{Analysis of the experiment}

\noindent The intensity of the scattered light field from a point particle 
is proportional to the term  $S = \cos^2(kx^\prime(t))\mathcal{F}(y,z)$. 
The experiment uses spheres of radius $R=200$ nm.
For the transverse motion, the sphere dimensions are small relative to the beam waist
$w=145$ $\mu$m so a point particle approximation is quite reasonable.
However, this is not the the case in the axial direction 
given the standing wave potential for $\lambda/2=532$ nm.
A previous study of the effect of the particle size on the optical coupling parameter $A$
\cite{Monteiro} showed that the behaviour $R=200$ nm remains qualitatively similar
to an point dipole particle. Nevertheless, for accuracy, here we convolve
the scattering function with a sphere of finite size in our simulations
and calculate for our scattering function:
 
\begin{equation}
 S(t) = \bar{S}(t)\mathcal{F}(y(t),z(t)),
\end{equation}
where:
\begin{equation}
\begin{aligned}
\bar{S}(t)&=\frac{1}{2}-\frac{3}{16(kR)^3}\cos(2kx^\prime (t))[2kR\cos(2kR)-\sin(2kR)]\\
&=\frac{1}{2}+C(kR)\cos(2kx^\prime (t)).
\label{finite}
\end{aligned}
\end{equation}

 The Fourier transform of the scattered field, yielding the frequency spectrum of scattered radiation,
allows us to extract the characteristic frequencies of the motions.
$\mathcal{F}(y(t),z(t))$ is modulated at the lowest frequencies (of order $100$ Hz) by the 
transverse secular  frequencies $\omega^{(y)}_{\textrm{S}}$ and  $\omega^{(z)}_{\textrm{S}}$; Fourier transforms of
traces of order several seconds show these clearly.

We focus now on Fourier transforms of the trapped regions, dominated by the mechanical and
drive frequencies. The scattered signal, in frequency space may be calculated using 
Eqs.\ref{ansatz} and \ref{finite}:

\begin{equation}
\bar{\textbf{S}}(\omega)=C(kR)\sum_{\textrm{(m,n)}}(-1)^lJ_{\textrm{m}}(2kX_{\textrm{M}}^{\infty})J_{\textrm{n}}(2kX_{\textrm{d}})\delta[\omega-(m\omega_{\textrm{M}} + n\omega_{\textrm{d}})],
\label{FT}
\end{equation}

\noindent where $m+n=2l,l\in Z$, while peaks with $m+n=2l+1$ are absent from the spectrum. The relative amplitude of the peaks $(m,n)$ and $(m^\prime,n^\prime)$ in the spectrum is: 

\begin{equation}
\mathcal{A}_{\textrm{rel}}=\frac{J_{\textrm{m}^\prime}(2kX_{\textrm{M}}^{\infty})J_{\textrm{n}^\prime}(2kX_{\textrm{d}})}{J_{\textrm{m}}(2kX_{\textrm{M}}^{\infty})J_{\textrm{n}}(2kX_{\textrm{d}})}.
\end{equation}

\noindent If  $X_{\textrm{M}}^{\infty}\ll X_d\lesssim(1/k)$ then $\mathcal{A}_{\textrm{rel}}\sim \displaystyle\frac{\Gamma(m+1)}{\Gamma(m^\prime+1)}(kX_{\textrm{M}}^{\infty})^{m^\prime-m}$, leading to an exponential reduction of higher order peaks (here $\Gamma$ is the Gamma function).\\

 Inspection of  Eq.\ref{FT} shows that the dominant terms in the spectrum
 involving the mechanical frequency are:

\begin{equation}
C(kR)^{-1} \bar{\textbf{S}}(\omega) \sim \bar{\textbf{S}}^+_{1}(\omega)+\bar{\textbf{S}}^-_{1}(\omega)+\bar{\textbf{S}}_{2}(\omega)
\end{equation}

where:
\begin{eqnarray}
\bar{\textbf{S}}^\pm_1(\omega)\simeq \mp J_1(2kX_{\textrm{M}}^{\infty})J_{\pm 1}(2kX_{\textrm{d}})\delta(\omega-\omega_{\textrm{M}}\pm\omega_{\textrm{d}})
\end{eqnarray}
and
\begin{equation}
\bar{\textbf{S}}_2(\omega)=-J_0(2kX_{\textrm{d}})J_2(2kX_{\textrm{M}}^{\infty})\delta(\omega-2\omega_{\textrm{M}}).
\end{equation}

The first terms $\bar{\textbf{S}}^\pm_1(\omega)$ predicts a pair of peaks of similar amplitude centred at frequencies $\omega_{\textrm{M}}\pm\omega_{\textrm{d}}$. The second term $\bar{\textbf{S}}_2(\omega)$
 produces a peak at the second harmonic
 (i.e. at double the mechanical frequency).
We find that both these features are a common feature of Fourier transforms in the trapped regime
although the experimental peaks are broadened due to the spread in $\omega_{\textrm{M}}$ as well as other thermal broadening effects.

An important point is that the relative amplitudes of these
peaks provide a means to estimate the steady state amplitudes 
$X_{\textrm{M}}^{\infty}$ and thus enable an estimate the equilibrium temperature of the nanosphere.
In particular, if there is strong cooling:

\begin{equation}
\left|\frac{J_0(2kX_{\textrm{d}})J_2(2kX_{\textrm{M}}^{\infty})}{J_{\pm 1}(2kX_{\textrm{d}})J_1(2kX_{\textrm{M}}^{\infty})}\right|
\sim \frac{X_{\textrm{M}}^{\infty}}{X_{\textrm{d}}} \to 0,
\end{equation} 
\noindent resulting in the suppression of the second harmonic ($2\omega_{\textrm{M}}$) peak in the spectrum.
Thus we expect that those traces for which the second peak is weak or nearly absent correspond
 to the strongest cooling. In this case, we estimate the reduction in temperature from
$\displaystyle\frac{T_{\textrm{eq}}}{T_{\textrm{int}}} \sim \left[\frac{X_{\textrm{M}}^{\infty}}{X_{\textrm{M}}(t=0)}\right]^2$
where $T_{\textrm{int}}$ corresponds to an initial temperature which we can choose to define either as the height of the optical well
or of a bath at $T=300$ K while $T_{\textrm{eq}}$ is our final equilibrium temperature.

\section{Optomechanical cooling rates $\Gamma_{\textrm{opt}}$}
In the regime of separable mechanical and drive motion (where 
Eq.\ref{ansatz} is a good approximation), the optomechanical damping may
be calculated by standard methods (e.g. Linear Response theory as applied to Eqs.\ref{xcool}
and \ref{acool}). 

Alternatively,  we note that Eqs.\ref{xcool}
and \ref{acool} are the equations of motion corresponding (in the semiclassical limit)
to a well-studied effective Hamiltonian which describes typical optomechanical
systems:

\begin{equation}
 \frac{{\hat H_{\textrm{lin}}}}{\hbar} =  -\Delta {\hat a}^\dagger {\hat a} 
  + \omega_M{\hat b}^\dagger {\hat b} + g ( {\hat a}+  {\hat a}^\dagger) ( {\hat b}+  {\hat b}^\dagger)/\sqrt{2}
\label{HLIN}
\end{equation}
where the coordinates have been rescaled ${\hat x} \to  \sqrt{2} X_{\textrm{ZPF}}{\hat x}$,
while ${\hat p} \to \sqrt{\hbar m\omega_{\textrm{M}}} {\hat p}$
and ${\hat x}=( {\hat b}+{\hat b}^\dagger)/\sqrt{2}$.
Here, $X_{\textrm{ZPF}} =\sqrt{\hbar/2m\omega_{\textrm{M}}}$ gives the scale of the oscillator ground state
(zero-point fluctuations).

The dynamics corresponding to Eq.\ref{HLIN} are extensively investigated
and the corresponding the optical damping $\Gamma_{\textrm{opt}}$ has been shown to be given by:
\begin{equation}
\Gamma_{\textrm{opt}}= g^2 \kappa\left[S(\omega_{\textrm{M}})-S(-\omega_{\textrm{M}})\right]
\label{cooling}
\end{equation}

\noindent where $S(\omega)=\displaystyle \left[(\Delta-\omega)^2 +\frac{\kappa^2}{4}\right]^{-1}$.

For the case of the levitated system investigated here, Eqs.\ref{xcool}
and \ref{acool} are fully equivalent to the Hamiltonian Eq.\ref{HLIN} provided we set
for the instantaneous optomechanical coupling:

\begin{equation}
 g^2(t)=\displaystyle\frac{\hbar k^2 A^2\bar{\alpha}^2}{m \omega_{\textrm{M}}}\sin^2(2kx_0(t))
\end{equation}
with $\omega_{\textrm{M}}^2 \simeq (2\hbar k^2 A \bar{\alpha}^2/m) \cos(2kx_0)\approx  2\hbar k^2 A \bar{\alpha}^2/m $ since $\omega_{\textrm{M}}$ is less sensitive to the excursion in $x_0$.

 Averaging over the period of the  micromotion $\displaystyle T_{\textrm{d}}=\frac{2\pi}{\omega_d}$:
\begin{equation}
\frac{1}{T_{\textrm{d}}}\int_0^{T_{\textrm{d}}}\cos(4kX_{\textrm{d}}\cos(\omega_{\textrm{d}} t))dt=\frac{1}{2\pi}\int_0^{2\pi}\cos(4kX_{\textrm{d}}\cos(u))du=J_0(4kX_{\textrm{d}}),
\end{equation}

\noindent we obtain:
\begin{equation}
\overline{g^2}=\frac{\hbar k^2 A^2\bar{\alpha}^2}{2m \omega_{\textrm{M}}}[1-J_0(4kX_{\textrm{d}})].
\end{equation}

The cooling rates are thus obtained by setting $ \overline{g} \equiv g$ in Eq.\ref{cooling}.

%%%%%%%%%%%%%%%%%%%%%%%%%%%%%%%%%%%%%%%%%%%%

% max 30 References

%%%%%%%%%%%%%%%%%%%%%%%%%%%%%%%%%%%%%%%%%%


\begin{thebibliography}{99}

\renewcommand{\familydefault}{\rmdefault}
\footnotesize 

% Force Sensing
\bibitem{Geraci10}
Geraci, A. A., Papp, S. B. and Kitching, J.
Short-Range Force Detection Using Optically Cooled Levitated Microspheres.
{\it Phys. Rev. Lett.} {\bf 105}, 101101 (2010)

% Gravitational Collapse
\bibitem{Diosi87}
Di\'{o}si, L.
A universal master equation for the gravitational violation of quantum mechanics.
{\it Phys. Lett. A} {\bf 120}, 377 (1987)

% Gravitational Collapse
\bibitem{Penrose96}
Penrose, R.
On gravity's role in quantum state reduction.
{\it Gen. Relativ. Gravit.} {\bf 28}, 581 (1996)

% Macroscopic superpositions
\bibitem{Arndt14}
Arndt, M. and Hornberger, K.
Testing the limits of quantum mechanical superpositions.
{\it Nature Phys.} {\bf 10}, 271 (2014)

% Single atom trapping
\bibitem{Grangier01}
Schlosser, N., Reymond, G., Protsenko, I., and Grangier, P.
Sub-poissonian loading of single atoms in a microscopic dipole trap.
{\it Nature} {\bf 411}, 1024 (2001)

% Cell trapping
\bibitem{Ashkin87}
Ashkin, A., Dziedzic, J. M., and Yamane, T.
Optical trapping and manipulation of single cells using infrared laser beams.
{\it Nature} {\bf 330}, 769 (1987)

% BEC review
\bibitem{Leggett01}
Leggett, A. J.
Bose-Einstein condensation in the alkali gases: Some fundamental concepts.
{\it Rev. Mod. Phys.} {\bf 73}, 307 (2001)

% Lattices review
\bibitem{Bloch05}
Bloch, I.
Ultracold quantum gases in optical lattices.
{\it Nature Phys.} {\bf 1}, 23 (2005)

% Raizen Brownian
\bibitem{Raizen10}
Li, T., Kheifets, S., Medellin, D. and Raizen, M. G.
Measurement of the Instantaneous Velocity of a Brownian Particle.
{\it Science} {\bf 328}, 1673 (2010)

% ICFO non-eq
\bibitem{Gieseler14}
Gieseler, J., Quidant, R., Dellago, C. and Novotny, L.
Dynamic relaxation of a levitated nanoparticle from a non-equilibrium steady state.
{\it Nature Nano.} {\bf 9}, 358 (2014)

% Our thermo paper
\bibitem{Millen14}
Millen, J., Deesuwan, T., Barker, P. F. and Anders, J.
Nanoscale temperature measurements using non-equilibrium Brownian dynamics of a levitated nanosphere.
{\it Nature Nano.} {\bf 9}, 425 (2014)

% Lehnert sideband cooling
\bibitem{Lehnert11}
Teufel, J. D. {\it et al.}
Sideband cooling of micromechanical motion to the quantum ground state.
{\it Nature} {\bf 475}, 359 (2011)

% Painter sideband cooling
\bibitem{Painter11_2}
Chan, J. {\it et al.}
Laser cooling of a nanomechanical oscillator into its quantum ground state.
{\it Nature} {\bf 478}, 89 (2011)

% Transduction
%\bibitem{Painter11}
%Safavi-Naeini, A. H. and Painter, O.
%Proposal for an optomechanical traveling wave phonon-photon translator.
%{\it New J. Phys.} {\bf 13}, 69 (2011)

% Cleland transduction
\bibitem{Cleland13}
Bochmann, J., Vainsencher, A., Awschalom, D. D. and Cleland, A. N.
Nanomechanical coupling between microwave and optical photons.
{\it Nature Phys.} {\bf 9}, 712 (2013)

% Lehnert & Regal transduction
\bibitem{Lehnert14}
Andrews, R. W. {\it et al.}
Bidirectional and efficient conversion between microwave and optical light.
{\it Nature Phys.} {\bf 10}, 321 (2014)

% Quantum network
\bibitem{Hakonen14}
Silanp{\"a}{\"a}, M. A. and Hakonen, P.
Optomechanics: Hardware for a quantum network.
{\it Nature} {\bf 507}, 45 (2014)

% Squeezing
\bibitem{Painter13}
Safavi-Naeini, A. H. {\it et al.}
Squeezed light from a silicon micromechanical resonator.
{\it Nature} {\bf 500}, 185 (2013)

% Two field cooling them
\bibitem{Chang10}
Chang, D. E. {\it et al.}
Cavity opto-mechanics using an optically levitated nanosphere.
{\it Proc. Natl Acad. Sci. USA} {\bf 107}, 1005 (2010)

% Feedback
\bibitem{Li11}
Li, T, Kheifets, S. and Raizen, M. G.
Millikelvin cooling of an optically trapped microsphere in vacuum.
{\it Nature Phys.} {\bf 7}, 527 (2011)

% Feedback
\bibitem{Gieseler12}
Gieseler, J., Deutsch, B., Quidant, R. and Novotny, L.
Subkelvin Parametric Feedback Cooling of a Laser-Trapped Nanoparticle.
{\it Phys. Rev. Lett.} {\bf 109}, 103603 (2012)

% Interferometry
\bibitem{Romero-Isart11}
Romero-Isart, O. {\it et al.},
Large Quantum Superpositions and Interference of Massive Nanometer-Sized Objects.
{\it Phys. Rev. Lett.} {\bf 107}, 020405 (2011)

% Cavity cooling proposal
\bibitem{Horak97}
Horak, P. {\it et al.}
Cavity-induced atom cooling in the strong coupling regime.
{\it Phys. Rev. Lett.} {\bf 79}, 4974 (1997)

% Cavity cooling 
\bibitem{Vuletic03}
Chan, H. W., Black, A. T. and Vuleti\'c, V.
Observation of collective-emission-induced cooling of atoms in an optical cavity.
{\it Phys. Rev. Lett.} {\bf 90}, 063003 (2003)

% Cavity cooling 
\bibitem{Rempi04}
Maunz, P. {\it et al.}
Cavity cooling of a single atom.
{\it Nature} {\bf 428}, 50 (2004)

% Cavity cooling an ion
\bibitem{Vuletic09}
Leibrandt, D. R., Labaziewicz, J., Vuleti\'c, V. and  Chuang, I. L.
Cavity Sideband Cooling of a Single Trapped Ion.
{\it Phys. Rev. Lett.} {\bf 103}, 103001 (2009)

% Peter cooling
\bibitem{Barker10}
Barker, P. F. and Shneider, M. N.
Cavity cooling of an optically trapped nanoparticle.
{\it Phys. Rev. A} {\bf 81}, 023826 (2010)

% Superpositions
\bibitem{Isart10}
Romero-Isart, O., Juan, M. L., Quidant, R. and Cirac, J. I.
Toward quantum superposition of living organisms.
{\it New J. Phys.} {\bf 12}, 033015 (2010)

% Arndt cooling
\bibitem{Asenbaum13}
Asenbaum, P. {\it et al.}
Cavity cooling of free silicon nanoparticles in high vacuum.
{\it Nature Commum.} {\bf 4}, 2743 (2013)

% Aspelmeyer cooling
\bibitem{Kiesel13}
Kiesel, N. {\it et al.}
Cavity cooling of an optically levitated submicron particle.
{\it Proc. Natl Acad. Sci. USA} {\bf 110}, 14180 (2013)

% Feedback
\bibitem{Gieseler13}
Gieseler, J., Novotny, L. and Quidant, R.
Thermal nonlinearities in a nanomechanical oscillator.
{\it Nature Phys.} {\bf 9}, 806 (2013)

% Ground state cooling us
\bibitem{Monteiro}
Monteiro, T. S. {\it et al.}
Dynamics of levitated nanospheres: towards the strong coupling regime.
{\it New J. Phys.} {\bf 15}, 015001 (2013)

\bibitem{March}
March, R. E., An Introduction to Quadrupole Ion Trap Mass Spectrome-
try, {\it J. Mass Spectrom.} \textbf{32}, 351 (1997)


% Cleland cooling
%\bibitem{Cleland10}
%O'Connell, A. D. {\it et al.},
%Quantum ground state and single phonon control of a mechanical resonator.
%{\it Nature} {\bf 464}, 697 (2010)

% Two field cooling us
%\bibitem{Pender12}
%Pender, G. A. T. {\it et al.},
%Optomechanical cooling of levitated spheres with doubly-resonant fields.
%{\it Phys. Rev. A} {\bf 85}, 021802 (2012)
























\end{thebibliography}
\end{document}